\documentclass[aps,prl,twocolumn,english,superscriptaddress]{revtex4}
\usepackage{graphicx}
\usepackage{amsmath}
\usepackage{amssymb}
\usepackage{amsfonts}
\usepackage{bm}
\usepackage{pifont}
\usepackage[colorlinks=true, linkcolor=blue, citecolor=blue, urlcolor=blue]{hyperref}

\newcommand{\MeV}{\textrm{MeV}}
\newcommand{\fm}{{\rm \, fm}}
\newcommand{\be}{\begin{equation}}
\newcommand{\ee}{\end{equation}}
\newcommand{\bea}{\begin{eqnarray}}
\newcommand{\eea}{\end{eqnarray}}
\newcommand{\pa}{\partial}

\newcommand{\cmark}{\ding{51}}
\newcommand{\xmark}{\ding{55}}

\newcommand{\figref}[1]{Figure~\ref{#1}}
\newcommand{\tabref}[1]{Table~\ref{#1}}

\begin{document}

\title{Skyrmions coupled to the sigma meson}

\author{Christoph Adam}
\affiliation{Departamento de F\'isica de Part\'iculas, Universidad de Santiago de Compostela and Instituto Galego de F\'isica de Altas Enerxias (IGFAE) E-15782 Santiago de Compostela, Spain}
\author{Carlos Naya}
\affiliation{Universidad de Alcalá, Departamento de Física y Matemáticas, 28805 Alcalá de Henares, Spain}
\author{Thomas Winyard}
\affiliation{Division of Mathematics, University of Dundee, Dundee DD1 4HN, United Kingdom}

\begin{abstract}
We propose a geometric coupling of the Skyrme model of nuclei to the scalar--isoscalar sigma meson by identifying the sigma field with the radial length of the chiral four-vector while preserving the topological protection of the baryon number. Restricting to the $B=1$ hedgehog ansatz, an asymptotic analysis of the field equations shows that the in-vacuum sigma mass $m_\sigma$ is replaced inside the nucleon by an effective in-medium mass $m_{\sigma,{\rm eff}} = 2m_\pi$. Full numerical integration via a double-shooting method and parameter optimization demonstrate that including the sigma meson leads to a remarkable improvement in the description of nucleon observables and to a pion decay constant very close to its physical value.
\end{abstract}

\maketitle

\section{Introduction}
In nuclear physics, the interaction between nucleons at long and intermediate distances is described in terms of meson exchange, including scalar isovector pions, vector isovector rho mesons, and the vector isoscalar omega meson, as realized in one-boson-exchange models, modern chiral effective field theory \cite{Weinberg,Epelbaum,Machleit}, and the hidden local symmetry approach \cite{Bando,Harada}. In particular, the strong intermediate-range attraction in the scalar--isoscalar channel is parameterized by the exchange of an effective scalar sigma ``meson,'' which phenomenologically represents correlated two-pion exchange and corresponds to the broad $f_0(500)$ resonance with a mass of order $500$--$600~\MeV$ \cite{Hatsuda,Dono,Kiselev}. A closely related description arises in relativistic mean-field approaches to nuclear matter, most notably in the Walecka model, where the dominant contributions to the nuclear interaction are provided by scalar sigma and vector omega mean fields \cite{Wal1,Wal2}.

The Skyrme model \cite{skyrme,HS-book,zahed-brown,weigel-book,NM-book,MR-book} is formulated in terms of pion fields, identifying nucleons and nuclei with topological soliton solutions (``skyrmions'') so that pion exchange is included by construction. 
The Skyrme model as originally introduced by Skyrme \cite{skyrme} contains no free parameters once the energy and length scales have been fixed. Even so, it provides a framework for the description of a wide range of properties of baryons, nuclei, and nuclear matter \cite{adkins1}--\cite{leask3}, some of which are difficult to describe in other frameworks (see \cite{NM-book} for a general discussion and, e.g., \cite{wood2,lau,halcrow,naya1} for more specific results).
The quantitative precision in the description of nuclear observables, on the other hand, is typically of the order of 20-30\% and, therefore, is currently not competitive with other methods of nuclear physics. But one obvious direction of improvement is the inclusion of additional mesons in line with other effective field theory (EFT) approaches.

Several proposals exist for coupling rho and omega mesons to the Skyrme model \cite{adkins3,sato,adkins4,gud2,hal,leask1,leask2}, or a tower of vector mesons via hidden local symmetry \cite{ma1,ma2} or holographic methods in flat space \cite{SutBPS1,SutBPS2,naya1}. The leading contribution of the omega meson in a derivative expansion can also be incorporated via the sextic term, which is sixth order in derivatives of the Skyrme field \cite{jackson1,Ding-Yan,BPS}.

By contrast, attempts to introduce the sigma meson into the Skyrme framework have been scarce. A generalization based on the linear sigma model \cite{Cohen} loses the topological protection of the baryon charge, whereas in Refs.~\cite{Schechter,Yabu,Schwes} the sigma meson was essentially identified with the dilaton field. Here we introduce a geometric proposal for coupling the sigma meson to the Skyrme model and work out its physical consequences for the $B=1$ hedgehog skyrmion.

In the linear sigma model, the sigma meson $\tilde{\sigma}$ and pions $\pi^a$ ($a=1,2,3$) form a chiral four-vector $\Phi^A = (\tilde{\sigma}, \pi^1, \pi^2, \pi^3)^t$ ($A=0,1,2,3$) on $\mathbb{R}^4$. Imposing $\Phi^A \Phi^A = f_\pi^2$, where $f_\pi$ is the pion decay constant, yields the nonlinear sigma model with field space topology $\mathbb{S}^3$, which is required to support topological skyrmions. From the Skyrme model perspective, it is therefore natural to reintroduce the sigma meson as the length of $\Phi^A$ \cite{chanfray},
\be
\Phi^A = (\Phi^0, \Phi^1, \Phi^2, \Phi^3)^t = \bar{\Sigma} (N^0, N^1, N^2, N^3)^t \, , 
\ee
where $N^A N^A = 1$.
The field space topology is $\mathbb{S}^3 \times \mathbb{R}_+$, supporting skyrmions as long as $\bar{\Sigma} = 0$ is avoided. Here $\bar{\Sigma}$ is a chiral singlet, while $N^A$ transforms nonlinearly and defines the $\mathrm{SU}(2)$ Skyrme field $U = N^0 \mathbf{1} + i \tau^a N^a$, with $\pi^a = f_\pi N^a$ and vacuum expectation value (VEV) $\langle \bar{\Sigma} \rangle = f_\pi$. Following standard Skyrme conventions, we use $F_\pi = 2f_\pi$ (with physical value $F_\pi = 186~\MeV$) and define the dimensionless field $\Sigma = (2/F_\pi)\bar{\Sigma}$ with $\langle \Sigma \rangle = 1$.

Our proposal is to replace $(F_\pi/2) N^A$ by $\Phi^A = \bar{\Sigma} N^A$ in those derivative terms of the Skyrme model coupled to the sigma meson, introducing no new unknown coupling constants. The potential terms, on the other hand, are chosen phenomenologically to reproduce the pion and sigma masses, fix $\langle \Sigma \rangle = 1$, and prevent unwinding at $\Sigma = 0$. We use the metric signature $(+---)$ and nuclear physics units ($\MeV$ and $\fm$) with $c=1$ and $\hbar = 197.3~\MeV\fm$.

\section{The Skyrme field coupled to the sigma meson}
We start from the generalized Skyrme model $\mathcal{L}^{(0)} = \mathcal{L}_2^{(0)} + \mathcal{L}_4^{(0)} + \mathcal{L}_6^{(0)} + \mathcal{L}_0^{(0)}$, where
\bea
\mathcal{L}_2^{(0)} &=& \frac{F_\pi^2}{8\hbar} \pa_\mu N^A \pa^\mu N^A , \\
\mathcal{L}_4^{(0)} &=& - \frac{\hbar}{8e^2} (\pa_\mu N^A \pa_\nu N^B - \pa_\nu N^A \pa_\mu N^B)^2 , \\
\mathcal{L}_6^{(0)} &=& -\Lambda^2 \pi^4 \mathcal{B}_\mu \mathcal{B}^\mu , \quad
\mathcal{L}_0^{(0)} = - \frac{m_\pi^2 F_\pi^2}{4\hbar^3} (1-N^0) ,
\eea
and $\mathcal{B}^\mu = \frac{1}{12\pi^2} \epsilon^{\mu\nu\rho\sigma} \epsilon^{ABCD} N^A \pa_\nu N^B \pa_\rho N^C \pa_\sigma N^D$ is the topological baryon current. Since $F_\pi$ is twice the VEV of $\bar{\Sigma}$, its explicit appearance indicates which terms should couple to $\Sigma$. Choosing a ``minimal'' coupling where only terms containing explicit factors of $F_\pi$ ($\mathcal{L}_2^{(0)}$ and $\mathcal{L}_0^{(0)}$) are coupled yields
\be
\mathcal{L} = \mathcal{L}_2 + \mathcal{L}_4 + \mathcal{L}_6 + \mathcal{L}_{0,1} + \mathcal{L}_{0,2} ,
\ee
with $\mathcal{L}_4 = \mathcal{L}_4^{(0)}$, $\mathcal{L}_6 = \mathcal{L}_6^{(0)}$, and
\bea
\mathcal{L}_2 &=& \frac{1}{2\hbar}\pa_\mu \Phi^A \pa^\mu \Phi^A = \frac{F_\pi^2}{8\hbar} \left( \pa_\mu \Sigma \pa^\mu \Sigma + \Sigma^2 \pa_\mu N^A \pa^\mu N^A \right) , \nonumber \\
\mathcal{L}_{0,1} &=& - V_\pi = -\frac{m_\pi^2 F_\pi^2}{4\hbar^3}\Sigma^2 (1-N^0) , \\
\mathcal{L}_{0,2} &=& - V_\Sigma = - \frac{m_\sigma^2 F_\pi^2}{32\hbar^3} \left( \Sigma^2 + \frac{1}{\Sigma^2} - 2 \right) . \nonumber
\eea
Coupling $\mathcal{L}_2$ automatically generates a canonical kinetic term for $\Sigma$. The potential $V_\Sigma$ provides the VEV $\langle \Sigma \rangle = 1$, diverges at $\Sigma = 0$ to prevent topological unwinding, and yields the sigma mass term $V_\Sigma \sim \frac{F_\pi^2}{8\hbar^3} m_\sigma^2 \sigma^2 + \mathcal{O}(\sigma^3)$ upon expanding $\Sigma = 1 - \sigma$. Note that $\mathcal{L}_6$ arises from integrating out the $\omega$ meson, with $\pi^4 \Lambda^2 = \hbar^3 g_\omega^2 / (2m_\omega^2)$, and therefore should not couple to $\Sigma$. We also leave $\mathcal{L}_4$ uncoupled for simplicity; replacing $N^A \to \Sigma N^A$ in {\em all} higher-derivative terms would multiply them by powers of $\Sigma$ and trigger a Derrick-type shrinking instability near $\Sigma \approx 0$ unless extra higher-derivative or nonminimal $\Sigma$ terms were added.

The physical values of the model parameters are (see e.g. \cite{NS-EoS} for $\Lambda^2$)
\bea \label{phys-val}
&F_\pi = 186\,\MeV, \quad m_\pi = 138\,\MeV,& \nonumber\\
&\Lambda^2 \sim 10\,\MeV\fm^3, \quad m_\sigma \sim 500\text{--}600\,\MeV.&
\eea
Introducing Skyrme units of length and energy,
\be \label{skyrme-units}
\bar{l} = \frac{2\hbar}{eF_\pi} \; , \quad \bar{E} = \frac{F_\pi}{4e} \, ,
\ee
and the dimensionless parameters
\be \label{model-const}
m = \frac{2m_\pi}{eF_\pi}\, , \quad \tilde{m} = \frac{2m_\sigma}{eF_\pi}\, , \quad
\lambda^2 = \frac{\Lambda^2 F_\pi^2 e^4}{2\hbar^3} \, ,
\ee
the static energy density reads $\mathcal{E} = (\bar{E}/\bar{l}^3)\,\tilde{\mathcal{E}}$, with $\tilde{\mathcal{E}} = \tilde{\mathcal{E}}_2 + \tilde{\mathcal{E}}_4 + \tilde{\mathcal{E}}_6 + \tilde{\mathcal{E}}_{0,1} + \tilde{\mathcal{E}}_{0,2}$ and
\bea
\tilde{\mathcal{E}}_2 &=& \pa_j \Sigma \pa_j \Sigma + \Sigma^2 \pa_j N^A \pa_j N^A , \nonumber \\
\tilde{\mathcal{E}}_4 &=& \frac{1}{2} \left[ (\pa_j N^A \pa_j N^A)^2 - (\pa_j N^A \pa_j N^B)^2 \right] , \nonumber \\
\tilde{\mathcal{E}}_6 &=& \pi^4 \lambda^2 (\mathcal{B}^0)^2 , \quad
\tilde{\mathcal{E}}_{0,1} = 2m^2 \Sigma^2 (1 - N^0) , \nonumber \\
\tilde{\mathcal{E}}_{0,2} &=& \frac{\tilde{m}^2}{4}\left( \Sigma^2 + \frac{1}{\Sigma^2} - 2 \right) .
\eea

\section{The hedgehog ansatz}
Inserting the $B=1$ hedgehog ansatz $N^0 = \cos f(r)$, $N^a = (x^a/r) \sin f(r)$, and $\Sigma = \Sigma(r)$ into the dimensionless energy $\tilde{E} = \int d^3x\,\tilde{\mathcal{E}} = 4\pi \int_0^\infty dr\,\mathcal{F}(r)$ gives $\mathcal{F} = \mathcal{F}_2 + \mathcal{F}_4 + \mathcal{F}_6 + \mathcal{F}_{0,1} + \mathcal{F}_{0,2}$, where
\bea
\mathcal{F}_2 &=& r^2 \Sigma'^2 + r^2 \Sigma^2 f'^2 + 2 \Sigma^2 \sin^2 f , \nonumber \\
\mathcal{F}_4 &=& 2f'^2 \sin^2 f + r^{-2} \sin^4 f , \quad
\mathcal{F}_6 = \frac{\lambda^2}{4r^2} \sin^4 f f'^2 , \nonumber \\
\mathcal{F}_{0,1} &=& 2 m^2 r^2 \Sigma^2 (1-\cos f) , \nonumber \\
\mathcal{F}_{0,2} &=& \frac{\tilde{m}^2}{4} r^2 (\Sigma^2 + \Sigma^{-2} - 2) .
\eea
The resulting Euler--Lagrange (EL) equations for $f(r)$ and $\Sigma(r)$ are
\bea
&&\left( r^2 \Sigma^2 + 2 \sin^2 f + \frac{\lambda^2}{4r^2} \sin^4 f \right) f'' \nonumber \\
&&+ \left( 2r\Sigma^2 + 2r^2 \Sigma \Sigma' - \frac{\lambda^2}{2r^3} \sin^4 f \right) f' \nonumber \\
&&+ \sin 2f \left( 1 + \frac{\lambda^2}{4r^2} \sin^2 f \right) f'^2 \nonumber \\
&&- \sin 2f \left( \Sigma^2 + \frac{\sin^2 f}{r^2} \right) - m^2 r^2 \Sigma^2 \sin f = 0, \label{EL-Sk}
\eea
and
\bea
&&r^2 \Sigma'' + 2r\Sigma' - \Sigma \left[ r^2 f'^2 + 2 \sin^2 f - 2 m^2 r^2 (1 - \cos f ) \right] \nonumber \\
&&- \frac{\tilde{m}^2}{4} r^2 \left( \Sigma - \Sigma^{-3} \right) = 0. \label{EL-sig}
\eea

For $r \to \infty$, the fields approach their vacuum values $f(r) \to 0$ and $\Sigma(r) = 1 - \sigma(r) \to 1$. In the linearized EL equation for $f$, $\sigma$ only enters nonlinearly, so $f$ satisfies the standard equation $r^2 f'' + 2rf' - (2 + m^2 r^2)f = 0$ with asymptotic solution
\be \label{f-large}
f_{\rm lin}(r) = \delta \left(\frac{1}{r^2} + \frac{m}{r}\right) e^{-mr} .
\ee
For $\sigma$, the homogeneous linearized equation ($f=0$), $r^2 \sigma'' + 2r\sigma' - \tilde{m}^2 r^2 \sigma = 0$, has the Yukawa solution $\sigma_{\rm h,lin} = (k/r) e^{-\tilde{m}r}$. In the presence of the hedgehog tail \eqref{f-large}, however, the inhomogeneous linearized equation to quadratic order in $f$ reads
\bea
r^2 \sigma'' + 2r\sigma' - \tilde{m}^2 r^2 \sigma &=& -\left( r^2 f'^2 + 2 f^2 - m^2 r^2 f^2\right) \nonumber \\
&=& -\delta^2 \Big( \frac{6}{r^4} + \frac{12m}{r^3} + \frac{11m^2}{r^2} \nonumber \\
&& + \frac{6m^3}{r} + 2 m^4 \Big) e^{-2mr} .
\eea
Since experimentally $\tilde{m} > 2m$, the $e^{-2mr}$ source tail dominates over the homogeneous $e^{-\tilde{m}r}$ tail at large $r$. The particular solution can be expanded as
\be \label{sigma-lin}
\sigma_{\rm lin}(r) = \delta^2 \left( \sum_{n=2}^\infty \frac{\chi_n}{r^n} \right) e^{-2mr} ,
\ee
where the coefficients $\chi_n$ are determined recursively; the first two are
\be \label{chi-coeff}
\chi_2 = \frac{2m^4}{\tilde{m}^2 - 4m^2} \, , \quad \chi_3 = \frac{6m^3 \tilde{m}^2 - 16 m^5}{(\tilde{m}^2 - 4 m^2)^2} .
\ee
The poles at $\tilde{m}=2m$ and the sign change of $\chi_2$ for $\tilde{m}<2m$ highlight the physical condition $\tilde{m} > 2m$. Crucially, the vacuum Yukawa tail governed by $\tilde{m}$ is replaced by an exponential decay governed by $2m$: inside the nucleon, the sigma meson acquires an effective in-medium mass $\tilde{m}_{\rm eff} = 2m$ (i.e., $m_{\sigma,{\rm eff}} = 2m_\pi$). This matches chiral effective theory and unitarized $\pi\pi$ results in nuclear matter, where the $f_0(500)$ pole shifts toward the $2m_\pi$ threshold with increasing density due to partial chiral restoration 
\cite{Hatsuda,HatsudaKunihiro1999,Chiang-Oset,GomezNicolaPelaez2002}.

Near the origin ($r \to 0$), regularity requires the Taylor expansions
\be \label{small-r-exp}
f(r) = \pi - b_1 r - b_3 r^3 - \dots \; , \quad
\Sigma(r) = s_0 + s_2 r^2 + \dots \, ,
\ee
where $b_1$ and $s_0 = \Sigma(0) \in (0,1)$ are free parameters, while higher coefficients follow recursively from Eqs.~\eqref{EL-Sk}--\eqref{EL-sig}:
\bea
s_2 &=& \frac{s_0}{24}\left[ 12b_1^2 - 16 m^2 + \tilde{m}^2\left(1-\frac{1}{s_0^{4}} \right) \right] , \nonumber \\
b_3 &=& \frac{b_1\left( \lambda^2 b_1^6 - 4 b_1^4 - 8 b_1^2 s_0^2 - 6 m^2 s_0^2 - 24 s_0 s_2\right)}{15( b_1^4 \lambda^2 + 8 b_1^2 + 4 s_0^2 )} .
\eea

\section{Numerical method and results}
For simplicity, we assume $\tilde{m} = 4m$ in the following, which corresponds to $m_\sigma = 4m_\pi = 552~\MeV$ and is right in the center of the expected interval.
To solve the coupled boundary value problem \eqref{EL-Sk}--\eqref{EL-sig} for a given parameter pair $(m,\lambda)$, we follow \cite{speightwin} and employ a double-shooting method. Using the small-$r$ expansion \eqref{small-r-exp} at $r_0 \ll 1$ and the combined asymptotic expansion at $r_\infty \gg 1$,
\bea
f(r) &\sim& \delta \left(\frac{m}{r} + \frac{1}{r^2}\right)e^{-mr}, \\
\Sigma(r) &\sim& 1 - \delta^2 \left(\frac{\chi_2}{r^2} + \frac{\chi_3}{r^3}\right)e^{-2mr} - \frac{k}{r} e^{-\tilde{m}r} ,
\eea
the boundary data are specified by four shooting parameters $\bm{c} = (b_1, s_0, \delta, k)$. We integrate the state vector $\bm{\Psi}(r) = (f, f', \Sigma, \Sigma')$ outward from $r_0$ to an intermediate matching radius $r_c$ and inward from $r_\infty$ to $r_c$. The residual $\bm{R}(\bm{c}) = \bm{\Psi}_{\rm L}(r_c) - \bm{\Psi}_{\rm R}(r_c)$ is driven to zero (within a tolerance of $10^{-8}$) via a Newton--Raphson flow $\bm{c}(t+\Delta t) = \bm{c}(t) - \Delta t\, J^{-1}\bm{R}$, with the Jacobian $J_{ij} = \pa R_i / \pa c_j$ evaluated by central finite differences. We also checked our method against an arrested Newton flow method \cite{spewin} with fixed boundary conditions given by the shooting.

\begin{figure}
\centerline{\includegraphics[width=\linewidth]{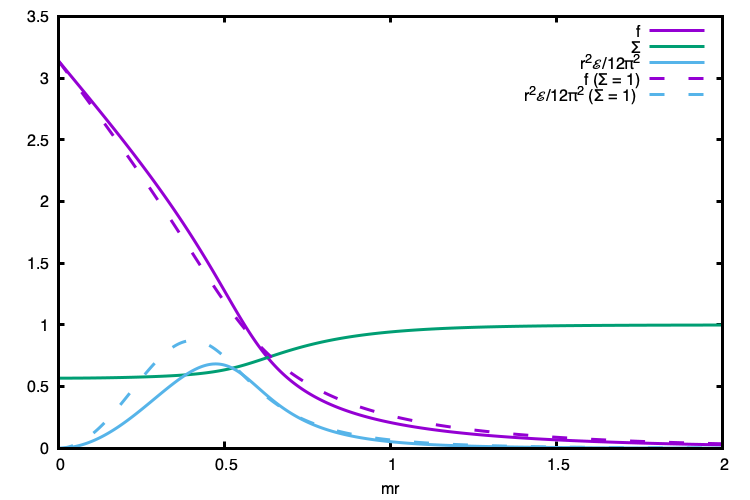}}
\caption{The solution to the field equations for ($m$, $\lambda$) that minimise $\Delta$ excluding $M_S$ (optimal fitting) given in \tabref{tab:results}, for the coupled model (solid) and uncoupled (dashed) cases. We include the hedgehog profile $f$, sigma meson $\Sigma$, and normalised radial energy density $\mathcal{E}r^2/12\pi^2$.}
\label{fig:field}
\end{figure}

\begin{table*}[t]
\centering
\caption{Fitted model parameters $(m,\lambda,\mu)$, cost function $\Delta$, and ratios of calculated nucleon observables to their empirical values (first row) with (\cmark) and without (\xmark) the sigma meson. An asterisk ($*$) indicates quantities included in the $n$-observable fit.}
\label{tab:results}
\begin{tabular}{c|c|c|c|c|c|c|c|c|c|c|c|c}
\hline\hline
$\Sigma$ & $m$ & $\lambda$ & $n$ & $\Delta$ & $\mu$  & $M_S$ & $I$ & $r_e$ & $r_m$ &  $g_{\pi NN}$ & $r_i$ & $\Lambda$ \\
\hline
-- & fit & fit & 8 & 0 & 1 & $866.7\,\MeV$ & $198.7\,\MeV\fm^2$ & $0.72\,\fm$ & $0.81\,\fm$ & 13.5 & $0.88\,\fm$ & $3.162\,\MeV^{1/2}\fm^{3/2}$ \\
\hline\hline
\cmark &  0.408 &  0 &  6 &  0.097 & *0.866 &  1.413 & *1.039 & *0.810 & *0.967 & *1.012 & *1.000 & 0 \\
\xmark &  0.415 &  0 &  6 &  0.129 & *0.792 &  1.429 & *0.987 & *0.775 & *0.978 & *1.065 & *1.020 & 0 \\
\hline
\cmark &  0.451 &  0 &  7 &  0.135 & *0.745 & *1.178 & *1.011 & *0.877 & *1.046 & *0.897 & *1.057 & 0 \\
\xmark &  0.464 &  0 &  7 &  0.157 & *0.678 & *1.180 & *0.971 & *0.851 & *1.067 & *0.942 & *1.081 & 0 \\
\hline
\cmark &  0.072 &  66.30 &  7 &  0.055 & *0.969 &  1.274 & *1.084 & *0.906 & *0.968 & *0.943 & *1.015 & *0.997 \\
\xmark &  0.011 &  2979.65 &  7 &  0.103 & *0.844 &  1.256 & *1.015 & *0.884 & *1.010 & *1.075 & *1.104 & *0.859 \\
\hline
\cmark &  0.055 &  125.26 &  8 &  0.085 & *0.887 & *1.134 & *1.050 & *0.947 & *1.010 & *0.861 & *1.045 & *1.017 \\
\xmark &  0.036 &  256.40 &  8 &  0.121 & *0.815 & *1.177 & *0.954 & *0.886 & *1.013 & *0.986 & *1.083 & *0.831 \\
\hline\hline
\end{tabular}
\end{table*}

From the converged profile for $(m,\lambda)$, we compute the nucleon observables within the rigid-rotor quantization \cite{adkins2}, where the nucleon ($j=1/2$) and $\Delta$ ($j=3/2$) masses are $M_j = M_S + \hbar^2 j(j+1)/(2I)$ in terms of the classical soliton mass $M_S$ and spin moment of inertia $I$. Using the experimental masses $M_N = 939~\MeV$ and $M_\Delta = 1232~\MeV$ and assuming that the spin contribution is the only quantum correction yields the empirical values $M_S^{\rm emp} = 866.7~\MeV$ and $I^{\rm emp} = 198.7~\MeV\fm^2$. We also evaluate the isoscalar electric and magnetic charge radii $r_e$ and $r_m$, the isovector electric charge radius $r_i$, the pion--nucleon coupling $g_{\pi NN}$, and the sextic coupling $\Lambda$.

Keeping the pion mass fixed at its physical value $m_\pi = 138~\MeV$, we allow the model pion decay constant to scale as $F_\pi = \mu F_{\pi,{\rm phys}}$ relative to its physical value $F_{\pi,{\rm phys}} = 186~\MeV$, as is common practice in the Skyrme model. Thus, $\mu$ serves both as a parameter and as an observable with physical target $\mu_{\rm phys} = 1$. Defining $e = 2m_\pi / (m F_\pi) = 1.484/(\mu m) \equiv e_0/\mu$ (with $e_0 = 1.484/m$), the physical observables are given in terms of the dimensionless functionals by
\bea
M_S &=& 31.33\,\mu^2 m \tilde{E}\,\MeV, \quad
I = 63.84\,\mu^2 m^3 R_I\,\MeV\fm^2, \nonumber\\
r_e &=& 1.427\,m\sqrt{R_e}\,\fm, \quad
r_m = 1.427\,m\sqrt{R_m/R_e}\,\fm, \nonumber\\
r_i &=& 1.427\,m\sqrt{R_i/R_I}\,\fm, \nonumber\\
g_{\pi NN} &=& \frac{16\pi M_N}{3e^2 F_\pi}\delta = \frac{16\pi \mu M_N}{3e_0^2 F_{\pi,{\rm phys}}}\delta, \nonumber\\
\Lambda &=& \frac{\sqrt{2\hbar^3}\,\lambda}{F_\pi e^2} = \frac{\sqrt{2\hbar^3}\,\mu\lambda}{F_{\pi,{\rm phys}} e_0^2},
\eea
where $M_N = M_S + 3\hbar^2/(8I)$, and the dimensionless integrals are
\bea
R_e &=& -\frac{2}{\pi}\int_0^\infty dr\,r^2 f'\sin^2 f, \nonumber\\
R_m &=& -\frac{2}{\pi}\int_0^\infty dr\,r^4 f'\sin^2 f, \nonumber\\
R_I &=& \int_0^\infty dr \rho_I (r) , \quad R_i = \int_0^\infty dr\, r^2 \rho_I (r) , \nonumber \\
\rho_I &\equiv & \frac{16\pi}{3}\sin^2 f \Big( r^2\Sigma^2 + r^2 f'^2 + \sin^2 f + \frac{\lambda^2}{4}\sin^2 f f'^2 \Big) .\nonumber\\
&
\eea

To optimize the parameters $(\mu, m, \lambda)$, we minimize the root-mean-square relative deviation (cost function)
\be
\Delta(\mu, m, \lambda) = \left[ \frac{1}{n}\sum_{i=1}^n \left(\frac{\phi_i(\mu,m,\lambda) - \phi_i^{\rm phys}}{\phi_i^{\rm phys}}\right)^2 \right]^{1/2} ,
\ee
over $n$ selected observables $\phi_i \in \{\mu, M_S, I, r_e, r_m, g_{\pi NN}, r_i, \Lambda\}$. The results of this optimization, comparing the model with (\cmark) and without (\xmark) the sigma meson, are presented in \tabref{tab:results}. The first row lists the physical or empirical values $\phi_i^{\rm phys}$, while all other rows display the ratios $\phi_i / \phi_i^{\rm phys}$ for the fitted configurations. We consider both the massive Skyrme model ($\lambda=0$, where $\Lambda=0$ is excluded from the fit) and the generalized model ($\lambda\neq 0$), and perform fits both excluding $M_S$ ($n=6,7$, since the classical mass $M_S$ is expected to receive negative one-loop Casimir corrections \cite{MeiWal}) and including $M_S$ ($n=7,8$). 

In all cases shown in \tabref{tab:results}, coupling the Skyrme field to the sigma meson significantly reduces the cost function $\Delta$ and brings $\mu$ much closer to unity. The improvement is particularly striking in the generalized model ($\lambda \neq 0$) when $M_S$ is not fitted ($n=7$): including the sigma meson cuts the cost function nearly in half ($\Delta = 0.055$ vs.\ $0.103$), yields $\mu = 0.969$ (within $3\%$ of the physical pion decay constant $F_\pi = 186~\MeV$), and reproduces all fitted nucleon radii, $I$, $g_{\pi NN}$, and $\Lambda$ within a few percent, which should be compared with the typical 30\% precision achieved by the standard Skyrme model \cite{adkins2}. Furthermore, the fit leads to a classical mass excess of $M_S / M_S^{\rm emp} = 1.274$, which is precisely of the expected magnitude, because the one-loop Casimir corrections should be suppressed by $N_C^{-1}$ relative to $M_S$, where $N_C=3$ is the number of colors. But even when $M_S$ is included in the fit ($n=8$), the sigma meson lowers $\Delta$ from $0.121$ to $0.085$ and increases $\mu$ from $0.815$ to $0.887$.

For the optimal case $\lambda \not= 0$, $n=7$ and for the corresponding fit values for $m$ and $\lambda$, we show the
resulting solution for both the coupled and uncoupled case in \figref{fig:field}, where the coupled sigma meson causes the energy density to shift radially outward from the origin. A similar behaviour was observed for all other solutions. 
\section{Summary and discussion}
We have introduced a geometric proposal for incorporating the scalar--isoscalar sigma meson $\Sigma$ into the Skyrme model, which is required for a realistic description of nuclear properties in any effective field theory. Concretely, we identified $\Sigma$ with the radial length of the chiral four-vector while preserving the topological protection of the baryon charge. In the $B=1$ hedgehog sector, the asymptotic field equations naturally reproduce the well-known nuclear medium effect that the vacuum sigma mass $m_\sigma \sim 4m_\pi$ is replaced inside the nucleon by an effective mass $m_{\sigma,{\rm eff}} = 2m_\pi$. Full numerical integration via double shooting and multiparameter optimization shows that including the sigma meson leads to a remarkable improvement of the simultaneous fit of proton and neutron observables as well as the pion decay constant, which are now all within a few percent of their physical values.

These results constitute a very strong vindication for the correctness of our proposal and 
underscore the importance of the sigma meson for a realistic description of nuclear properties.  
At the same time, they open many avenues for future work like, e.g.:
(i) extending the $B=1$ analysis to include explicit vector mesons ($\rho, \omega$), to achieve an even more precise description of nucleon observables;
(ii) applying the skyrmion interaction formalism of Refs.~\cite{Hal1,Hal2} to study how the sigma field modifies the intermediate-range nucleon--nucleon interaction. In particular, $\Sigma$ could lead to an attractive central isoscalar nucleon-nucleon potential, which is difficult to achieve in the standard Skyrme model;
(iii) computing multiskyrmion solutions ($B>1$) to investigate clustering and binding energies in light nuclei \cite{naya1}; and
(iv) studying Skyrme crystals for dense nuclear matter, where the attractive sigma field is expected to soften the equation of state near saturation density and reduce the nuclear compression modulus \cite{Sk-crystal,leask2}, thereby resolving one of the most important shortcomings of the Skyrme model approach to nuclear matter.

{\em Note aded}: While finishing this manuscript, we were informed of a related work using a slightly different approach, where nucleon observables in the Skyrme model are calculated by identifying the sigma meson with the dilaton \cite{zwicky}.

\section*{Acknowledgements}
CA acknowledges financial support from the Spanish Research State Agency under project PID2023-152762NB-I00, the Xunta de Galicia under the project ED431F 2023/10 and the CIGUS Network of Research Centres, the María de Maeztu grant CEX2023-001318-M funded by MICIU/AEI /10.13039/501100011033, and the European Union ERDF. The authors thank the organisers of the conference nonlinear phenomena in soliton dynamics in Salamanca where the project was initially discussed.

\end{document}